# Timescale effect estimation in time-series studies of air pollution and health: A Singular Spectrum Analysis approach

**Massimo Bilancia and Girolamo Stea**

*Dipartimento di Scienze Statistiche "Carlo Cecchi"*
*Università degli Studi di Bari*
*e-mail:* `mabil@dss.uniba.it`; `stea@dss.uniba.it`

**Abstract:** A wealth of epidemiological data suggests an association between mortality/morbidity from pulmonary and cardiovascular adverse events and air pollution, but uncertainty remains as to the extent implied by those associations although the abundance of the data. In this paper we describe an SSA (Singular Spectrum Analysis) based approach in order to decompose the time-series of particulate matter concentration into a set of exposure variables, each one representing a different timescale. We implement our methodology to investigate both acute and long-term effects of $PM_{10}$ exposure on morbidity from respiratory causes within the urban area of Bari, Italy.

## Contents

*This work is partially supported by a grant of the University of Bari - Italy.
†Contacts: Massimo Bilancia, Department of Statistical Sciences "Carlo Cecchi", University of Bari - Via C. Rosalba n.53, 70124, Bari (Italy). Tel: +39(0)805049341, Fax +39(0)805049147, Email: mabil@dss.uniba.it

## 1. Introduction

A wealth of epidemiological studies based on time-series analysis has shown evidence for association between morbidity/mortality caused by respiratory and cardiovascular adverse events and the exposure to airborne particles [3]. Suspended total particulate matter is a complex mixture of organic and inorganic substances in either liquid or solid phase. They can vary in size, composition and origin and can be characterized both physically and chemically. Particles with an aerodynamic diameter of less than 10 microns are referred to as $PM_{10}$: they may be inhaled reaching the upper airways and the lungs, with risk for adverse effects on health.

Assuming counts data $y_t$ of daily adverse health event being distributed as conditionally independent Poisson given the rate $\varphi_t$, a standard ecological Poisson regression model (which has been used, with minor variations, in most of large scale epidemiological studies [11]) is

$$\log(\varphi_t) = \beta_0 + \beta_1 \mathrm{PM}_{10,t} + [\mathrm{DOW}_t] + \mathcal{S}(t, \delta_1) + \mathcal{S}(temp_t, \delta_2) + \mathcal{S}(umr_t, \delta_3) \quad (1.1)$$

where $\mathrm{PM}_{10,t}$ is measured in $\mu g/m^3$, $[\mathrm{DOW}_t]$ is a six-dimensional vector of dummy variables pointing the day of the week, and $\mathcal{S}(t, \delta_1)$ is a smooth term function of calendar time controlling for seasonality and other trends (the degree of roughness being controlled by the smoothing parameter $\delta_1$); further smooth confounders include temperature (temp) measured in °C and relative humidity (umr) expressed as percentage (meteorological confounders may affect the pollution-morbidity association: see [26] for an interesting discussion).

Among the most recent results we may cite the MISA 2, a planned study over 15 Italian cities for the period 1996–2002 [4]. Updated city-specific estimates show an overall RR=1.005 (C.I.: 0.991-1.018 - estimate $\pm 2$ std. err.) per 10 $\mu g/m^3$ increase in $PM_{10}$ concentration for respiratory causes, with a similar RR=1.005 (C.I: 1.000-1.010) for cardiovascular causes. The estimated lag-3 days overall RR of hospitalization due to respiratory causes is 1.006 (C.I.: 1.002-1.011), while RR=1.003 (C.I.: 1.000-1.006) is estimated for cardiovascular adverse events. Another important and recent study is the European meta-analysis conducted by the Regional Office of World Health Organization (WHO) for Europe, based on 17 country-specific estimates [1]; the mortality RRs reported in the WHO-meta analysis are 1.009 (C.I.: 1.005-1.013) for cardiovascular causes and 1.013 (C.I.: 1.005-1.020) for respiratory ones (per 10 $\mu g/m^3$ increase in $PM_{10}$).

Despite this growing body of evidence, a considerable uncertainty remains to be seen: this begs the question of whether these associations represent premature mortality within only few days among those already near to death. Such a displacement (or *harvesting*) effect has been discussed by several authors after [19], and can complicate the interpretation of the results: a reasonable underlying hypothesis is that mortality/morbidity displacement is related to associations on shorter time scales, while longer time scales are supposed to be resistant to mortality displacement. If associations reflect only harvesting, from a public-health point of view, the effect of air pollution on morbidity can be considered as having a limited impact. A first attempt to assess short and long-term effects was proposed in [28] by using Cleveland STL decomposition by means of LOESS smoothing algorithm to separate the time-series of daily deaths into long, intermediate and short (residual) timescale series. Similarly, [10] developed a methodology based on the Discrete Fourier Transform to obtain a set of orthogonal predictors at given timescales, by partitioning the base interval $[0, \pi]$ into a given set of Fourier frequencies. Expanding time series of particulate matter concentration into a set of lagged exposure variables is a concurrent approach: a distributed-lag model was proposed in [36] by replacing the pollutant effect with a distributed lag-specification, each lag coefficient representing a specific contribution to exposure: cumulative effects on a given timescale can be obtained by summing up contributions for a given range of lags.

The aforementioned methods share a common drawback: suitable timescales need to be arranged by the researcher in advance, rather than being a natural result of the data analysis process. For example, [28] examines mid-scale components of the daily number of deaths with smoothing windows of 15, 30, 45 and 60 days: risk estimates are provided for each mid-scale window without attempting to provide a data based criterion to choose among diverse alternatives. Similarly, [10] estimate the association between air pollution and mortality using six fixed timescale: $\geq 60$ days, 30–59 days, 14–29 days, 7–13 days, 3.5–6 days and $< 3.5$ days. In a word, current approaches to mortality displacement estimation do not provide automatic, data-driven methods to decompose air pollution time series into a set of suitable exposure variables, each one representing a different timescale. Improvements in this field would be greatly beneficial in public health time series studies. For this reason, we propose an alternative approach based on Singular Spectrum Analysis - SSA [14]. The word "spectrum" may be quite confusing here, since SSA is not derived from Fourier analysis, but it is an algorithmic technique rooted in dynamical system theory, linear algebra and multivariate geometry. SSA can be defined as a model-free approach to decompose a time series in easy-to-interpret components such as trend, harmonic intermediate components and pure noise (short scale residual). This task can be accomplished by exploiting a functional clustering algorithm based on a suitable metrics that allows a sensible grouping of more "elementary" components. No fixed timescales need to be known in advance in our novel approach: the proposed methodology is used to test the harvesting hypothesis on a dataset of residents in the city of Bari (Apulia, Italy).

The paper is structured as follows. Sec. 2 briefly reviews the data and the pre-processing methods used to deal with missing information and outliers; Sec. 3 gives a short introduction to SSA; the first part of Sec. 4 develops a functional clustering algorithm to group elementary components into interpretable exposure variables at several timescales; the second part of Sec. 4 describes timescale estimation by means of GAM models with integrated smoothing parameter selection, we reported also some results about our data set. Finally, Sec. 5 gives a brief discussion about the results and outlines future research opportunities.

## 2. Data description and pre-processing

Epidemiological data were obtained from Apulian Regional Epidemiological Center, concerning the daily time-series of hospitalized people among residents in the city of Bari between June 1th, 2000 and December 31th, 2001 (in total $N = 579$ days), diagnosed as suffering from pulmonary diseases (ICD-IX Classification: 460-519). Time series of particulate matter concentration and meteorological data were collected by a monitoring network subgroup of the Municipality of Bari (Department of Environmental Protection and Health), including four monitoring stations named "S. Nicola", "King", "Savoia", "Cavour" that collected information concerning a wide set of pollutants, such as Benzene, CO, NO, $NO_2$, $NO_x$, $O_3$, and $SO_2$. It is worth noticing that most time-series present a large number of missing data points.

Bi-hourly measurements of $PM_{10}$ were available on each of the four monitoring stations, whereas temperature and relative humidity were available on an hourly basis on "S. Nicola", "Savoia" and "Cavour" stations only. These data have been pre-processed in order to calculate an overall daily series for each variable (from midnight to midnight the day after). Preliminary data analysis concerned of the adjustment for most disturbing outliers attributable to temporary failures in monitoring devices. In fact, we computed a robust estimate of the covariance matrix of each multivariate time-series by using the Minimum Covariance Determinant (MCD) estimator (see [25] for details: the robust MCD estimator requires much less data for reliable results than the classical covariance matrix estimator, and gives more interpretable results as extreme values are well isolated.). We set an empirical rule by removing the five multivariate observations that showed the highest Mahalanobis distance (from the barycentre) based on MCD estimate. Fig. 1 shows some details for the $PM_{10}$ series : most of the bi-hourly data are considered as to be missing, as the MCD estimation algorithm removes forcibly all the rows containing at least one missing datum.

For each monitoring station we recovered daily measures by averaging the twenty-four hourly observations (the mean of the twelve two-hours observations in the case of $PM_{10}$) when at least 75% of one-hour observations were available (this criterion is compliant with APHEA protocol, [2]); otherwise, the daily datum was considered missing. After averaging over time, there were still a lot of missing values in the $PM_{10}$ series, as it is shown in the right panel of Fig. 2. The left panel contains the daily time series of hospitalizations.

Some exploratory statistics obtained after daily averaging are reported in Tab. 1: it is readily apparent that missing data in the $PM_{10}$ series is about 60% in the "King" station. Comparable values were observed for the other monitoring stations; temperature and relative humidity were far less difficult to analyze. In order to obtain an overall daily measure for each variable, we computed a spatial average of each daily time-series. This synthesis of information is mainly based on the assumption of constant exposure over the whole urban area. This is a reasonable assumption for temperature and relative humidity, whereas for the daily $PM_{10}$ series this can be justified on the ground that correlation coefficients between the measurements ranged from 0.66 to 0.87.

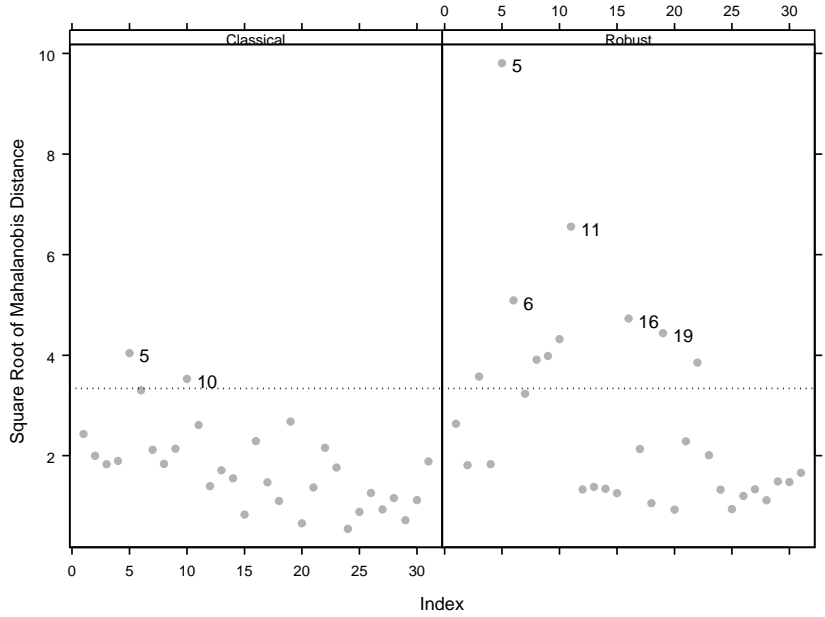

Fig 1. *LEFT - Classic squared Mahalanobis distance for each multivariate observation of the $PM_{10}$ hourly series. RIGHT - Robust Square Mahalanobis distance calculated through the output of the MCD algorithm (Minimum Covariance Determinant Estimator).*

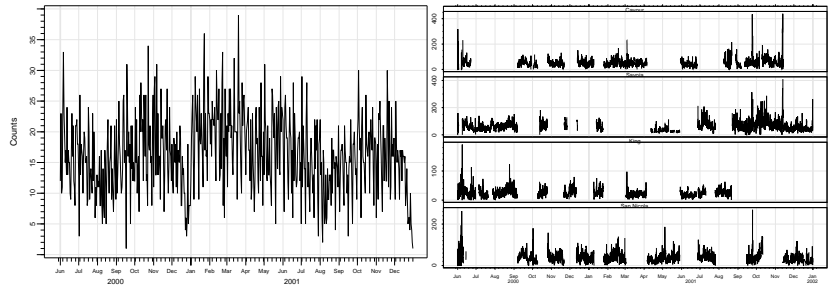

Fig 2. *LEFT - Time-series of hospitalized people between June 1th, 2000 and December 31th, 2001, diagnosed as suffering from pulmonary diseases (ICD-IX Classification: 460-519). RIGHT - Daily time series of $PM_{10}$ collected at the stations "Cavour", "Savoia","King" and "San Nicola".*

Some synthetic values of the reconstructed series can be found in Tab. 2; it is worth noticing the dramatic reduction in number of the missing points in the $PM_{10}$ series. There were still a few missing values in correspondence of those days for which the four $PM_{10}$ measurements were not available. In this case, 15-days causal moving averages was used for imputing missing values. The ultimate result of this information filtering and retrieval work is shown in Fig. 3.

Table 1
*Some exploratory statistics for $PM_{10}$, temperature (temp) and relative humidity daily (umr) time-series*

| Variables | Days | Missing | % Missing | Min | Median | Max |
|---|---|---|---|---|---|---|
| **$PM_{10}$** | | | | | | |
| S. Nicola | 579.00 | 304.00 | 52.5 | 8.195 | 35.431 | 94.083 |
| King | 579.00 | 350.00 | 60.45 | 9.354 | 23.345 | 54.637 |
| Savoia | 579.00 | 270.00 | 46.63 | 21.945 | 59.769 | 226.767 |
| Cavour | 579.00 | 321.00 | 55.44 | 6.954 | 47.186 | 280.388 |
| **temp** | | | | | | |
| S. Nicola | 579.00 | 86.00 | 14.8 | 2.065 | 19.481 | 34.565 |
| Savoia | 579.00 | 15.00 | 2.6 | 0.159 | 19.356 | 35.411 |
| Cavour | 579.00 | 20.00 | 3.45 | 2.537 | 20.549 | 34.546 |
| **umr** | | | | | | |
| S. Nicola | 579.00 | 114.00 | 19.68 | 19.955 | 64.844 | 97.752 |
| Savoia | 579.00 | 15.00 | 2.6 | 30.456 | 67.672 | 95.46 |
| Cavour | 579.00 | 27.00 | 4.66 | 17.22 | 66.588 | 80.264 |

Table 2
*Some exploratory statistics for the spatially aggregated $PM_{10}$, temperature (temp) and relative humidity daily (umr) time series*

| Variables | Days | Missing | % Missing | Min | Median | Max |
|---|---|---|---|---|---|---|
| $PM_{10}$ | 579.00 | 12.00 | 2.07 | 13.907 | 40.46 | 253.58 |
| temp | 579.00 | 0.00 | 0 | 1.587 | 19.202 | 34.841 |
| umr | 579.00 | 0.00 | 0 | 27 | 67.2 | 94.25 |

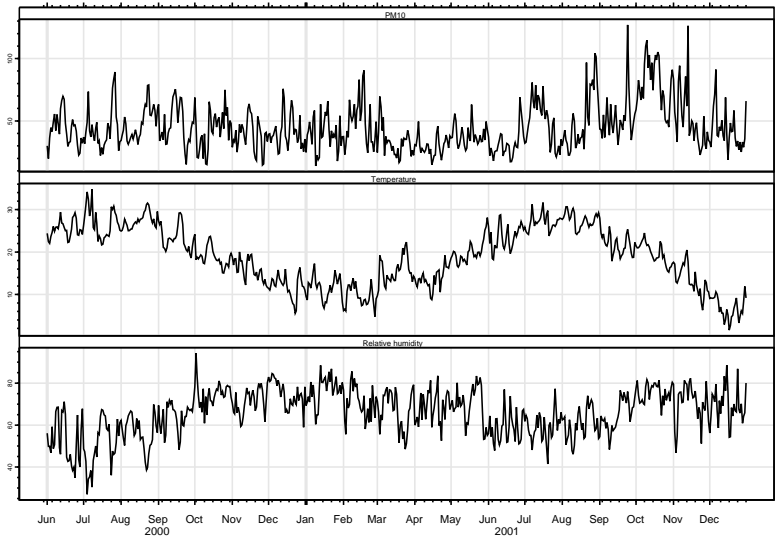

Fig 3. *Reconstructed $PM_{10}$, temperature and relative humidity time-series.*

## 3. Singular Spectrum Analysis

In this section a brief review of Singular Spectrum Analysis (SSA) is provided: detailed description can be found in [13; 14]. We denote the daily $PM_{10,t}$ time-series by $\{x_t\}_{t=0}^{N-1}$: SSA relies on the Karhunen-Loève decomposition of a covariance matrix estimate of $K$ lagged copies of the original time series [29; 34; 35].

Let $L < [N/2]$ be a fixed integer called *window length* and introduce $K = N-L+1$ *lagged vectors* $(x_{i-1}, x_{i-2}, \ldots, x_{i+L-2})^T$, for $i = 1, \ldots, k$: the first step consists of embedding the original time series $\{x_t\}_{t=0}^{N-1}$ into a lower-dimensional manifold than the original phase space, by transforming it into the following *L-trajectory matrix*

$$X = \begin{pmatrix} x_0 & x_1 & \ldots & x_{K-1} \\ x_1 & x_2 & \ldots & x_K \\ x_2 & x_3 & \ldots & x_{K+1} \\ \vdots & \vdots & \ddots & \vdots \\ x_{L-1} & x_L & \ldots & x_{N-1} \end{pmatrix}$$

Some celebrated results in dynamic system theory (a good reference is [5]) ensure that all the qualitative (topological) characteristics of the phase space are preserved after such a dimensionality reduction process. It is worth noticing that every L-trajectory matrix is Hankel, i.e. its entries coincide along the secondary matrix diagonals such that $i + j = s$ for $2 \leq s \leq L + K$, and vice versa every $L \times K$ Hankel matrix is the L-trajectory matrix of some time series.

In the second stage, further information collapsing is carried out by computing the eigenvalues $\lambda_i$ of the $L \times L$ matrix $S = XX^T$. Let $d = \text{rank}(X) = \text{rank}(XX^T)$ be the number of nonzero eigenvalues of the matrix $S$: if $d < L$ then $\lambda_1 \geq \lambda_2 \geq \ldots \geq \lambda_d > 0$ and $\lambda_{d+1} = 0$ for all other eigenvalues with indexes larger than $d$. The trajectory matrix can be decomposed into its Singular Value Decomposition (SVD, [12])

$$X = X_1 + \ldots + X_i + \ldots + X_d \tag{3.1}$$

where $X_i = \sqrt{\lambda_i} u_i v_i^T$, $\sqrt{\lambda_1} \geq \ldots \geq \sqrt{\lambda_d} > 0$ are the *singular values* of $S$, $u_i \in R^L$ ($i = 1, \ldots, d$) is a system of orthonormal eigenvectors associated to nonzero eigenvalues of $X$ and $v_i = X^T u_i / \sqrt{\lambda_i}$. Hence, the trajectory matrix is decomposed into a sum of elementary rank-one, pairwise bi-orthogonal matrices. The collection $(\sqrt{\lambda_i}, u_i, v_i)$ will be referred to as the *ith eigentriple* of the SVD (3.1).

In the third phase, SSA attempts to reconstruct components $\{x_{\ell t}\}_{t=0}^{N-1}$ such that the original time series can be decomposed into the sum of $p$ time-series

$$x_t = \sum_{\ell=1}^{p} x_{\ell t}, \qquad t = 0, \ldots, N-1 \tag{3.2}$$

which can have meaningful interpretations. Reconstruction of such components requires a suitable grouping of the set of indexes $I = \{1, \ldots, d\}$ into $p$ disjoint subsets $I_1, \ldots, I_\ell, \ldots, I_p$ with $I_\ell = \{\ell_1, \ldots, \ell_{n_\ell}\}$, such that the SVD expansion can be reformulated as

$$X = X_{I_1} + \ldots + X_{I_\ell} + \ldots + X_{I_p} \qquad (3.3)$$

where $X_{I_\ell} = X_{\ell_1} + \ldots + X_{\ell_{n_\ell}}$. If all the matrices on the right hand-side of (3.3) are Hankel, then they are L-trajectory matrices from which components $\{x_{\ell t}\}_{t=0}^{N-1}$ on different timescales can be easily reconstructed.

Nevertheless, this situation rarely occurs in practice: in most real data sets no sensible grouping can be found such that $X_{I_1}, \ldots, X_{I_p}$ are L-trajectory matrices. Last SSA phase consists of *Hankelization* or *diagonal averaging* of resultant matrices $X_{I_1}, \ldots, X_{I_p}$. Let $Y$ be any $L \times K$ matrix with elements $y_{ij}$, $L^\star = \min(L, K)$, $K^\star = \max(L, K)$, and $||Y||_\mathcal{M}^2 = \sum_{i=1}^{L} \sum_{j=1}^{K} y_{ij}^2$ the squared Frobenius-Perron norm of the matrix $Y$. Hankelization is carried out by means of a linear operator $\mathcal{H}$ acting on the space of $L \times K$ matrices: the resulting matrix $Z = \mathcal{H}Y$ with elements $z_{ij}$ is defined as follows ($s = i + j$)

$$z_{ij} = \begin{cases} \dfrac{1}{s-1} \sum_{j=1}^{s-1} y^\star_{j,s-j} & \text{for } 2 \leq s \leq L^\star \\ \dfrac{1}{L^\star} \sum_{j=1}^{L^\star} y^\star_{j,s-j} & \text{for } L^\star + 1 \leq s \leq K^\star + 1 \\ \dfrac{1}{N-s+2} \sum_{j=s-K^\star}^{L^\star} y^\star_{j,s-j} & \text{for } K^\star + 2 \leq s \leq N + 1 \end{cases} \qquad (3.4)$$

where $y^\star_{ij} = y_{ij}$ if $L < K$ and $y^\star_{ij} = y_{ji}$ otherwise. It can be proved that $Z$ is Hankel and thus it is the L-trajectory matrix of some time series. It is also the best approximation to $Y$ in the sense of Frobenius-Perron norm [6]: if $\mathcal{M}_{L \times K}$ is the space of real $L \times K$ matrices, and $\mathcal{M}^{(\mathcal{H})}_{L \times K}$ is the linear subspace of Hankel $L \times K$ matrices, then $||Y - Z||^2$ is minimum, so that it readily follows that $\mathcal{H} : \mathcal{M}_{L \times K} \to \mathcal{M}^{(\mathcal{H})}_{L \times K}$ is an orthogonal projector operator and $\mathcal{H}X = X$. By applying diagonal averaging to both members of (3.3) every resultant matrix $X_{I_1}, \ldots, X_{I_p}$ produces an L-trajectory matrix, from which the decomposition (3.2) can be easily recovered.

## 4. Eigentriple grouping

### *4.1. Reconstruction of components*

We can consider the *lag-covariance* matrix $C = S/K$ instead of $S$ for obtaining the singular values. $C$ is a sort of non-centered covariance matrix among columns of $X$ (the L-lagged vectors), and its use is justified by the fact that from $Cu_i = \lambda_i^C u_i$ it follows that $Su_i = \left(K\lambda_i^C\right)u_i$. The latter expression shows that the orthonormal system $\{u_i\}$ is unaffected by the choice of $C$, and the only difference in the SVD of $X$ lies in the magnitude of the corresponding eigenvalues (they are $K$ times larger for $S$). This fact simplifies the visual mining of component series. We set $L = 60$ for the window length, as adverse effects at timescales longer than two months are likely to be confounded with long-term effects due to other causes.

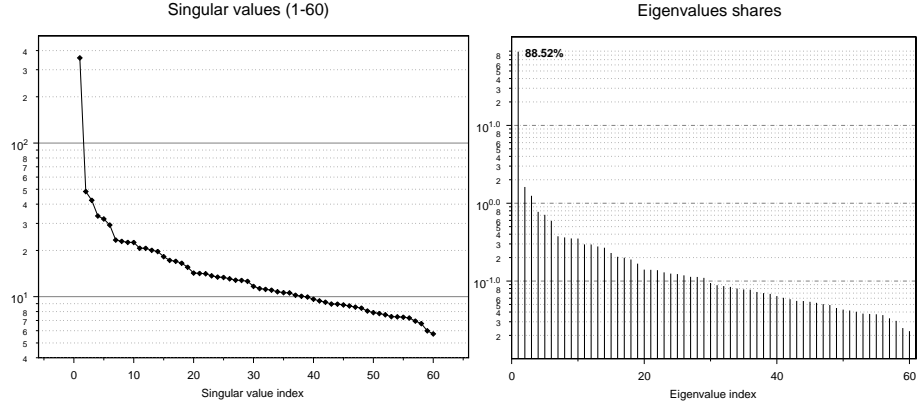

FIG 4. *LEFT - Singular values for eigentriples (1-60). RIGHT - Eigenvalues shares for eigentriples (1-60). For both graphics a semi-logarithmic scale on the Y axis has been used.*

The left panel of Fig. 4 shows spectrum of each singular value. Singular values are very close to each other, except for a large dominant value which prompts for a long-period slow-varying component (trend), and several plateaux that correspond to shorter period oscillatory components or pure noise. The right panel shows the degree of approximation of each single component of the SVD of $X$: it can be proved that $||X||^2_{\mathcal{M}} = \sum_{i=1}^{p} \lambda_i$ and that $X - X_{I_\ell}$ for $I_\ell = \{\ell_1, \ldots, \ell_{n_\ell}\} \subseteq \{1, \ldots, d\}$ is the SVD of $X_{I_J}$ with $I_J = \{1, \ldots, d\} / I_\ell$, hence a measure of the degree of approximation referred to as *eigenvalues share of the eigentriples with indexes* $I_\ell = \{\ell_1, \ldots, \ell_{n_\ell}\}$ can be defined in the following way

$$1 - \frac{||X - X_{I_\ell}||^2_{\mathcal{M}}}{||X||^2_{\mathcal{M}}} = \frac{\sum_{i=1}^{p} \lambda_i - \sum_{j \in \{1,\ldots,d\}/I_\ell} \lambda_j}{\lambda_1 + \ldots + \lambda_d} = \frac{\lambda_{\ell_1} + \ldots + \lambda_{\ell_{n_\ell}}}{\lambda_1 + \ldots + \lambda_d}$$

The largest eigenvalue ($\sqrt{\lambda_1} = 357.85$) accounts for about 88% of the total variability: this fact strengthens the belief that the corresponding eigentriple can be identified with slowly-varying component (trend), whereas the individuation of further components demands for a more elaborate algorithm.

A suitable decomposition of the $PM_{10}$ series can be determined by modifying the four-step algorithm described in the previous section. We suggest to apply Hankelization to each term in the full SVD decomposition (3.1): if $X_i^{(\mathcal{H})} = \mathcal{H} X_i$ then

$$X = X_1^{(\mathcal{H})} + \ldots + X_i^{(\mathcal{H})} + \ldots + X_L^{(\mathcal{H})} \qquad (4.1)$$

with $p = d = L = 60$, being $S = XX^T$ non singular. In general, elementary hankelized matrices on the right hand side of (4.1) are not pairwise bi-orthogonal (row and column orthogonal) matrices, hence the sum of two of such matrices does not need to be Hankel. It can be easily proved that $X_\ell^{(\mathcal{H})}$ and $X_\jmath^{(\mathcal{H})}$ are bi-orthogonal if and only if $\langle X_\ell^{(\mathcal{H})}, X_\jmath^{(\mathcal{H})} \rangle_\mathcal{M} = 0$, where $\langle X_\ell^{(\mathcal{H})}, X_\jmath^{(\mathcal{H})} \rangle_\mathcal{M} = \sum_{i=1}^{L} \sum_{j=1}^{L} x_{ij}$ is the inner-product compatible with the Frobenius-Perron norm. Additionally (see [14], page. 258), it can be proved that since $\langle X_\ell^{(\mathcal{H})}, X_\jmath^{(\mathcal{H})} \rangle_\mathcal{M} = 0$, $X_\ell^{(\mathcal{H})} + X_\jmath^{(\mathcal{H})}$ is an Hankel matrix. Hence it is the L-trajectory matrix of some component time series: this condition will be referred to as *weak L-separability*.

By joining elementary hankelized components having minimum distance in terms of weak L-separability, we often obtain a sensible grouping (3.3) whose component matrices are as close as possible (in the sense of the Frobenius-Perron distance) to Hankel matrices, hence amenable to a suitable interpretation after the diagonal averaging step. A sensible measure of weak L-separability between components $\ell$ and $\jmath$ is the *w-correlation*

$$w_{\ell\jmath}^{(\mathcal{M})} = \frac{\langle X_\ell^{(\mathcal{H})}, X_\jmath^{(\mathcal{H})} \rangle_\mathcal{M}}{||X_\ell^{(\mathcal{H})}||_\mathcal{M} ||X_\jmath^{(\mathcal{H})}||_\mathcal{M}}$$

where $\ell, \jmath = 1, \ldots, L$ and $||X_\ell^{(\mathcal{H})}||_\mathcal{M} = (\langle X_\ell^{(\mathcal{H})}, X_\ell^{(\mathcal{H})} \rangle_\mathcal{M})^{1/2}$

If the absolute value of the $w$-correlations is small then the corresponding series are almost $w$-orthogonal, but if $w_{\ell\jmath}^{(\mathcal{M})}$ is large then the two series are far from being $w$-orthogonal and therefore badly separable. If the matrix $W = \{|w_{\ell\jmath}^{(\mathcal{M})}|\}$ has a quasi-block diagonal structure, eigentriple grouping can be done accordingly by joining elementary components belonging to the same block. Unfortunately, real datasets usually have an entangled structure, except for some theoretical examples of little importance: Fig. 5 shows a 100-gray level representation of the $W$ matrix for the $PM_{10}$ series. Darkest gray on the main diagonal corresponds to $|w_{\ell\ell}^{(\mathcal{M})}| = 1$: a sharp block structure apparently stands for groups $\{1\}$ and $\{2-6\}$ only.

To overcome this difficulty we define the matrix $1 - W = \{1 - |w_{\ell\jmath}^{(\mathcal{M})}|\}$. From a formal point of view $1 - W$ is a dissimilarity matrix: it is symmetric and off-diagonal elements are strictly positive. Similar components have small $w$-correlations (in absolute value): equivalently, they have large values in the corresponding entries of the dissimilarity matrix $1 - W$. Therefore, it is natural to group elementary components by clustering hierarchically the elementary hankelized series taking $1 - W$ as the distance matrix. Consequently, the complete linkage method seems to be the best choice.

Results are shown in the right panel of Fig. 5: almost $p = 5$ groups are apparent (the choice of $p = 3$ and $p = 4$ is not compatible with the dendrogram branching). According to the clustering output, the following decomposition into $p = 5$ groups can be done: $\mathcal{G}_1 = \{1-6\}$, $\mathcal{G}_2 = \{7-23\}$, $\mathcal{G}_3 = \{24-33, 35, 36\}$, $\mathcal{G}_4 = \{34, 37-44\}$, $\mathcal{G}_5 = \{45-60\}$. Elementary components in (3.1) are grouped accordingly, and transformed into Hankel matrices to obtain the L-trajectory matrices of the corresponding reconstructed series.

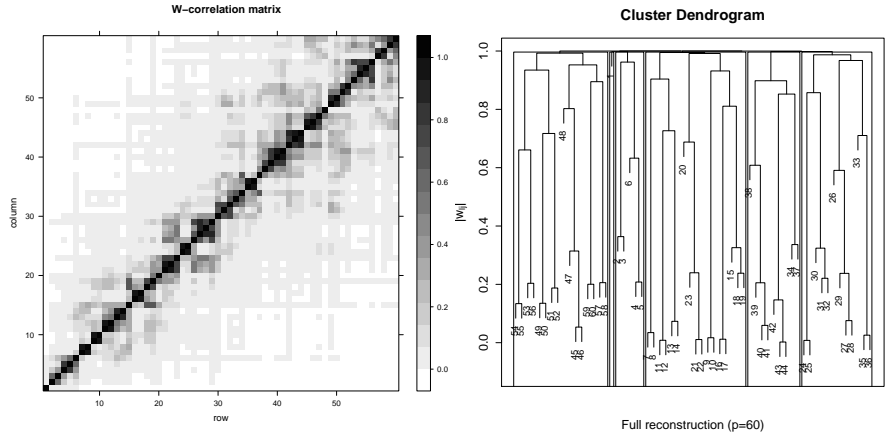

FIG 5. *LEFT - 100 gray-level representation of W matrix for hankelized elementary components of the SVD of $PM_{10}$ time-series. RIGHT - Hierarchical clustering of elementary hankelized series with $1 - W$ as dissimilarity matrix. Complete linkage method is used: $p=5$ groups are highlighted.*

Periods of the reconstructed components are not fixed in advance: therefore, the periodogram analysis of each reconstructed component may help us to estimate the corresponding timescales. More sophisticated approaches include non-parametric or parametric ARMA-based spectral density estimation [7; 8], which are both based on a suitable smoothing of the Fast Fourier Transform output, and require considerable skill for a proper application. For this reason, we suggest a faster and heuristic approach to period estimation. Period is defined as the amount of time necessary to complete a cycle: hence the number of turning peaks (local maxima) observed during the $N = 579$ days will be a rough estimate of the frequency (the number of cycles in the unit time). According to our definition, an approximate period estimate is given by the inverse of this quantity

$$\Pi\left(\mathcal{G}_i\right) = \frac{\text{Number of days in the observation period}}{\text{Number of peaks in the } i\text{th reconstructed component}}$$

Exploiting this simple estimator we found $\Pi\left(\mathcal{G}_1\right) \simeq 27.57$, $\Pi\left(\mathcal{G}_2\right) \simeq 7.24$, $\Pi\left(\mathcal{G}_3\right) \simeq 3.97$, $\Pi\left(\mathcal{G}_4\right) \simeq 2.84$, $\Pi\left(\mathcal{G}_5\right) \simeq 2.46$: "day" is the natural measurement unit.

It is not surprising that some redundancies may arise: for example, estimated timescales $\Pi\left(\mathcal{G}_4\right)$ and $\Pi\left(\mathcal{G}_5\right)$ are almost identical. This result can be explained by noting that there does not exist a simple one-by-one correspondence between weak L-separability and the frequency content of the elementary series which are joined together by our functional clustering algorithm. For this reason, if $[\Pi\left(\mathcal{G}_\ell\right)] = [\Pi\left(\mathcal{G}_j\right)]$ we shall say that reconstructed components $\ell$ and $j$ are *non-identifiable* (by $[\cdot]$ we denote the integer part function): in this case we prescribe that the corresponding elementary series should be merged into a single group, and a new component should be reconstructed by diagonal averaging.

This criterion seems to be reasonable for the application at hand, but it may be inappropriate in other contexts. For example, an orthogonal design is based on a set of orthogonal exposure variables: unfortunately $w$-orthogonality does not imply classical (Euclidean) orthogonality, with the result that reconstructed components are not pairwise orthogonal. In fact only elementary components in the right-hand side of (3.1) have this property, being the principal components of the L-trajectory matrix. In this case, a suitable additional constraint to remove non-identifiability would be $\left|R_{\mathcal{G}_\ell, \mathcal{G}_j}\right| < \varepsilon$, which consists in merging reconstructed series $\ell$ and $j$ for which the Pearson correlation coefficient $R_{\mathcal{G}_\ell, \mathcal{G}_j}$ is below a predetermined tolerance $\varepsilon$.

In our case components $\mathcal{G}_4$ and $\mathcal{G}_5$ are non-identifiable according to our definition and $R_{\mathcal{G}_4, \mathcal{G}_5} \simeq 0.19$. Therefore, a new grouping into $p = 4$ components can be obtained by merging elementary series formerly classified into groups $\mathcal{G}_4$ and $\mathcal{G}_5$: $\mathcal{G}_{new.1} = \mathcal{G}_1$, $\mathcal{G}_{new.2} = \mathcal{G}_2$, $\mathcal{G}_{new.3} = \mathcal{G}_3$ and $\mathcal{G}_{new.4} = \{34, 37-60\}$. The new reconstructed components are shown in Fig. 6: looking at the first two panels, a remarkable overfitting of the longest period waveform is apparent. We did not explore this feature, as prediction or change-point detection are not the objectives of our analysis. We propose the following interpretation of the estimated series:

- $\mathcal{G}_{new.1}$: eigenvalues $\{1 - 6\}$ are grouped. Given that $\Pi\left(\mathcal{G}_{new.1}\right) \simeq 27.57$, this variable measures particulate matter exposure at a timescale of about four weeks;
- $\mathcal{G}_{new.2}$: eigenvalues $\{7 - 23\}$ are grouped. This series can be considered as a proxy for exposures occurred in the past week, as $\Pi\left(\mathcal{G}_{new.2}\right) \simeq 7.24$;
- $\mathcal{G}_{new.3}$: eigenvalues $\{24 - 33, 35, 36\}$ are grouped. The reconstructed series is a short-period predictor, measuring lagged exposure at a five-day timescale: $\Pi\left(\mathcal{G}_{new.3}\right) \simeq 3.97$;
- $\mathcal{G}_{new.4}$: eigenvalues $\{34, 37 - 60\}$ are grouped. Very short periods of about three days or less are mixed up, as $\Pi\left(\mathcal{G}_{new.4}\right) \simeq 2.63$.

Of course, a main feature of SSA is that it encompasses an automated data-driven approach for decomposing the time series into timescale components: reconstructed short-period series can be used for testing the mortality displacement hypothesis. In addition, long-period air pollution effects can be estimated in correspondence of exposure variables that can be easily interpretable, and that vary at timescales of scientific interest (such as seasonality and trend). We address this issue in the next section.

### *4.2. Timescale estimation*

A GAM model accounting for the exposure variables at given timescales can be formulated as

$$\log\left(\varphi_t\right) = h_t + \sum_{\ell=1}^{p} \beta_\ell x_{\ell t} \qquad (4.2)$$

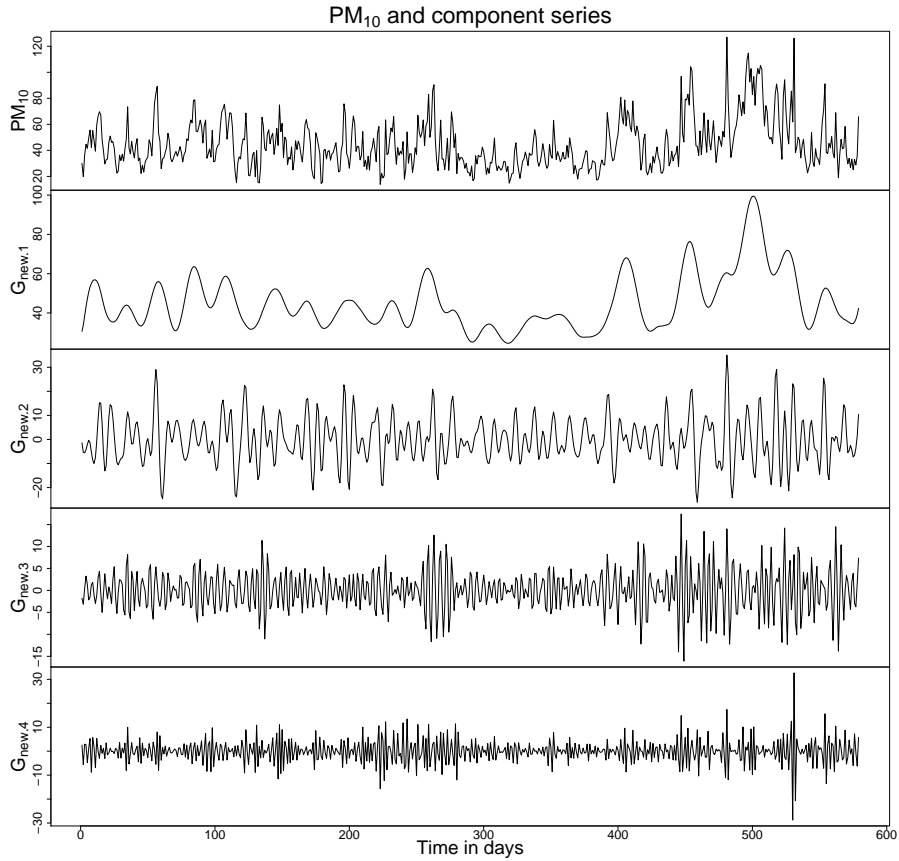

FIG 6. *Original $PM_{10}$ and $p = 4$ reconstructed components according to the following grouping: $\mathcal{G}_{new.1} = \{1-6\}$, $\mathcal{G}_{new.2} = \{7-23\}$, $\mathcal{G}_{new.3} = \{24-33, 35, 36\}$, $\mathcal{G}_{new.4} = \{34, 37-60\}$.*

where $h_t$ is a shortcut for the intercept, the dummy vector $[\text{DOW}_t]$ and the smooth components, while $\{x_{\ell t}\}_{\ell=0}^{N-1}$ is a suitable set of exposure variables chosen to account for timescale effects. Our functional clustering method can be considered as a dimensionality reduction algorithm that allows for parametrization of model (4.2) in term of a small number of waveforms $x_{\ell t}$; other methods, such as standard principal components analysis (PCA) and independent component analysis (ICA, [20]) assume mathematically convenient constraints (orthogonality and independence) but have no meaningful justification as they cannot decompose the original data into a set of exposure variables and complicate the interpretation of components.

The initial SSA decomposition can be modified on the ground of a careful consideration of both estimated timescales and singular value amplitudes (see

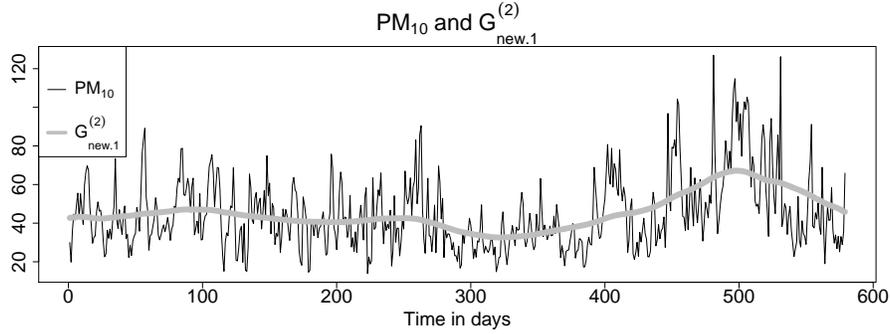

FIG 7. *Original $PM_{10}$ series and estimated trend on the basis of the new grouping strategy: $\mathcal{G}^{(2)}_{new.1} = \{1\}$ and $\mathcal{G}^{(2)}_{new.2} = \{2-6\}$.*

Fig. 4). For the first component we have $\Pi(\mathcal{G}_{new.1}) \simeq 27.57$, a timescale which is relatively shorter than the window length ($L = 60$). Intuitively, the first group includes too many elementary components, with the result that lower frequency harmonics have been smoothed away. It should be noted that a single leading singular value is apparent from Fig. 4: therefore, a new decomposition can be obtained by splitting group 1 into the two sub-groups $\{1\}$ and $\{2-6\}$. The new decomposition is given by: $\mathcal{G}^{(2)}_{new.1} = \{1\}$, $\mathcal{G}^{(2)}_{new.2} = \{2-6\}$, $\mathcal{G}^{(2)}_{new.3} = \mathcal{G}_{new.2}$, $\mathcal{G}^{(2)}_{new.4} = \mathcal{G}_{new.3}$, $\mathcal{G}^{(2)}_{new.5} = \mathcal{G}_{new.4}$. We have $\Pi(\mathcal{G}^{(2)}_{new.1}) = 144.75$ and $\Pi(\mathcal{G}^{(2)}_{new.2}) \simeq 27.57$ days: Fig. 7 shows the new grouping slow-varying component (trend) and original series.

Of course, other answers may be sensible: a trend plus season solution can be estimated by widening $\mathcal{G}^{(2)}_{new.1}$ in a suitable way. The estimated timescales of elementary components entering group $\mathcal{G}^{(2)}_{new.2} = \{2-6\}$ are $\Pi(2) \simeq 64.33$, $\Pi(3) = 48.25, \Pi(4) \simeq 32.17, \Pi(5) \simeq 26.32, \Pi(6) \simeq 22.26$. Therefore, it is logical to join the first and second leading singular values to obtain the following new decomposition: $\mathcal{G}^{(3)}_{new.1} = \{1-2\}$, $\mathcal{G}^{(3)}_{new.2} = \{3-6\}$, $\mathcal{G}^{(3)}_{new.3} = \mathcal{G}_{new.2}$, $\mathcal{G}^{(3)}_{new.4} = \mathcal{G}_{new.3}$, $\mathcal{G}^{(3)}_{new.5} = \mathcal{G}_{new.4}$, with $\Pi(\mathcal{G}^{(3)}_{new.1}) \simeq 72.38$ and $\Pi(\mathcal{G}^{(3)}_{new.2}) \simeq 27.57$.

Although the overall number of suitable alternative decompositions is not large, a subjective adjustment seems to be still required to obtain a correct and easy-to-interpret decomposition. A simple procedure for forward traversing the space of all sensible decomposition can rely on estimating the statistical goodness of fit of each set of exposure variables entering the GAM model (4.2), after adjusting for the increasing complexity by means of a suitable penalty: data-driven model choice is certainly appealing since the interest is focused on the relationship between air pollution and morbidity. A computationally feasible solution is the UBRE criterion - a rescaled version of the AIC statistics, see [16], pg. 160 - based on the minimization of the expected mean squared error (details are discussed in [32] and [33]): for $n$ independent observations from an exponential family with scale parameter $\phi$ (in the Poisson case $\phi = 1$) the UBRE has the following expression

$$\text{UBRE} = \frac{1}{n} \sum_{\ell=1}^{n} D\left(y_\ell; \hat{\mu}_\ell\right) + \frac{2\text{tr}\left(R\right)\phi}{n} - \phi$$

where $\sum_{\ell=1}^{n} D\left(y_\ell; \hat{\mu}_\ell\right)$ is the log-likelihood ratio statistics (Deviance) for the current model, $R$ is the weighted linear operator that produces estimates of the adjusted dependent variable at each step of the GAM local scoring estimation algorithm, and $\text{tr}\left(R\right)$ represents the overall effective degrees of freedom of the model (see [32] for details). It can be proved that UBRE is a simplified version of a generalized cross-validation (GCV) criterion that is valid when the scale parameter is not known; in addition, it is very similar to the GCV-$PM_{10}$ criterion introduced in [22].

Another benefit of automatic model selection concerns the choice of degrees of freedom (dfs), associated with smooth terms entering model (4.2) to adjust for temporal unmeasured confounders and meteorological variables. Each smooth term is a *natural cubic spline*, which can be derived from a conditional minimization problem with respect to coefficients of a suitable function basis, given a quadratic penalty depending on a symmetric smoothing matrix $S\left(\delta_j\right)$ [17]: for *fixed* values of the smoothing parameters $\delta_j$, parameter estimation of the GAM model (4.2) is easily shown to be equivalent to a conditional minimization problem with multiple quadratic penalties [33]. It is usually difficult to decide the amount of smoothing that is allowed for smooth functions in model (4.2): fixing each df to a pre-specified value may be sometimes justified on the ground of biological knowledge and of the problem peculiarities, although it is likely to invalidate the reproducibility of epidemiological findings [23]. For example [9] estimate a model very similar to (1.1) (except for the exclusion of the relative humidity term) by setting $\delta_1 = 7$, $\delta_2 = 8$: an identical model is estimated in [3] by choosing $\delta_1 = 7 \times \text{years}^{-1}$ (i.e. $\delta_1 = 3.5$ for a two years observation period) and $\delta_2 = 6$. We know that data-driven model choice is a necessary task in particulate matters time-series studies, even if the number of candidate models may be prohibitive: when $k$ smoothing terms are allowed into the model (as in our case), simultaneous fixed parameter estimation and smoothing parameter selection based on UBRE minimization over a $k$-dimensional space were computationally infeasible until the methods reported in [30; 31] were recently introduced. The related software, the R `mgcv::gam` V.1.3-29 library has been used for model estimation and selection throughout this section.

### 4.3. Results and some comparisons

To test the harvesting hypothesis we assumed that model (4.2) holds: evidence for short-term effects may reflect increased recruitment into the frail pool caused by air pollution (and not by a former disease condition): Table 3 shows effect estimates and approximate p-values for each predictor entering a model based on the $\mathcal{G}_{new}$ decomposition. The estimated degrees of freedom of each smooth term (edf) are inversely related to the smoothing parameter estimates: in order for a GAM to be identifiable each smooth evaluated at its covariate values should sum to zero ([16] refers to this as 'centering' the smoothing). This identifiability constraint removes one degree of freedom, thus the inferior limit for any smooth which is not curved at all (a straight line) is 1 (at variance with bounds on degrees of freedom for free smoothing splines, ranging between 2, a straight line, and $+\infty$, a perfectly interpolating spline).

TABLE 3
*Effect estimates, approximate significance of smooth terms and global model scores: the decomposition of $PM_{10}$ used here is $\mathcal{G}_{new.1} = \{1-6\}$, $\mathcal{G}_{new.2} = \{7-23\}$, $\mathcal{G}_{new.3} = \{24-33, 35, 36\}$, $\mathcal{G}_{new.4} = \{34, 37-60\}$*

| Approximate significance of parameter estimates | | | | |
|---|---|---|---|---|
| | Estimate | Std. Error | z value | $Pr(>|z|)$ |
| (Intercept) | 2.1852714 | 0.0614917 | 35.538 | <2e-16 |
| pm10.1.new | 0.0002876 | 0.0011013 | 0.261 | 0.794 |
| pm10.2.new | -0.0012147 | 0.0012534 | -0.969 | 0.332 |
| pm10.3.new | 0.0018025 | 0.0024178 | -0.746 | 0.456 |
| pm10.4.new | 0.0008980 | 0.0019772 | 0.454 | 0.650 |

| Approximate significance of smooth terms | | | |
|---|---|---|---|
| | edf | Chi.sq | p-value |
| S(day) | 8.879 | 186.168 | <2e-16 |
| S(temp) | 2.352 | 10.407 | 0.0645 |
| S(umr) | 1.976 | 5.668 | 0.2253 |

| Global scores | | | |
|---|---|---|---|
| R-sq.(adj) | Dev. explained | UBRE score | n |
| 0.501 | 51.6% | 0.36048 | 579 |

Conditionally on $\mathcal{G}_{new}$ we found little evidence for mortality displacement, as well as there were no associations present on longer timescales. We investigated the sensitivity of our results with respect to the chosen decomposition: the model based on the $\mathcal{G}_{new}^{(2)}$ set provided comparable global scores (UBRE=0.3618 and deviance explained equal to 51.6%). The only significant effect occurred at $\mathcal{G}_{new.1}^{(2)}$ (adjusted RR=1.02 with C.I.: 1.001-1.039), although a four-month association between morbidity and air pollution is quite unrealistic and it may be likely due to data-snooping.

When $\mathcal{G}_{new}^{(3)}$ was used, global model scores showed a significant decrease (approximate Anova tests comparing the current model with the previous ones were both significant). Results are shown in Table 4: the smooth term controlling for unmeasured temporal confounders was significant, as well as significant effects occurred at $\mathcal{G}_{new.1}^{(3)}$ and $\mathcal{G}_{new.2}^{(3)}$ timescales. These results suggest to reject the harvesting hypothesis: the negative effect estimated at the $\mathcal{G}_{new.1}^{(3)}$ timescale (corresponding to a relative-risk decrease) may be associated with a pool of healthy individuals which are still healthy two months after exposure to air pollution. The largest effect occurred at the timescale of one month (RR=1.00, C.I.=1.001-1.007), which corresponds to a relative-risk increase of about 4% per 10 $\mu g/m^3$ increase of $PM_{10}$ concentration. These findings are quite comparable to those of [10], which reported (by means of their FFT-based decomposition method applied to both cardiovascular and respiratory causes in four US cities) larger effects at longer timescales from 14 days up to 2 months.

TABLE 4
*Effect estimates, approximate significance of smooth terms and global model scores: the decomposition of $PM_{10}$ used here is $\mathcal{G}^{(3)}_{new.1} = \{1-2\}$, $\mathcal{G}^{(3)}_{new.2} = \{3-6\}$, $\mathcal{G}^{(3)}_{new.3} = \{7-23\}$, $\mathcal{G}^{(3)}_{new.4} = \{24-33, 35, 36\}$, $\mathcal{G}^{(3)}_{new.5} = \{34, 37-60\}$*

| Approximate significance of parameter estimates | | | | |
|---|---|---|---|---|
| | Estimate | Std. Error | z value | $Pr(>|z|)$ |
| (Intercept) | 2.5895118 | 0.1390330 | 18.625 | <2e-16 |
| pm10.1.new(2) | -0.0087046 | 0.0029872 | -2.914 | 0.00357 |
| pm10.2.new(2) | 0.0039257 | 0.0016050 | 2.446 | 0.01445 |
| pm10.3.new(2) | -0.0011658 | 0.0012576 | -0.927 | 0.35392 |
| pm10.4.new(2) | 0.0019135 | 0.0024188 | 0.791 | 0.42889 |
| pm10.5.new(2) | 0.0009902 | 0.0019600 | 0.505 | 0.61341 |
| Approximate significance of smooth terms | | | | |
| | edf | Chi.sq | p-value | |
| S(day) | 8.936 | 166.976 | <2e-16 | |
| S(temp) | 3.006 | 13.403 | 0.0629 | |
| S(umr) | 2.016 | 6.807 | 0.2354 | |
| Global scores | | | | |
| | R-sq.(adj) | Dev. explained | UBRE score | n |
| | 0.508 | 52.4% | 0.34438 | 579 |

We applied the Dominici's methodology to our $PM_{10}$ data for a direct comparison: suitable R software (the decompose() function described in [10]) was used to generate two new sets of exposure variables. For the first one, the "breaks" in the decompose function have been set to $1, 19, 41, 83, 165, 579$ days: except for the first and the last, these values are defined as $579/r$ where $r = 30, 14, 7, 3.5$. This choice was used as a standard in [10], and it generated quite equivalent waveforms to those obtained by the SSA $\mathcal{G}^{(3)}_{new}$-based decomposition. According to [10] the interpretation of such timescales is: more than 30 days (timescale 1, trend plus large scale periodicity, see Fig. 8), 14–30 days (timescale 2) and so on down to 1–3.5 days (timescale 5). A second decomposition was generated by setting breaks equal to $1, 21, 80, 146, 220, 579$; except for the first and the last these were obtained by taking $r = 27.57, 7.24, 3.97, 2.63$. The three decompositions (in a word SSA, FFT-A and FFT-B) are shown side by side in Fig. 8: surprisingly, no significant effects occurred conditionally on FFT-A (see Table 5), an impractical result which was confirmed by the poor global model score performance (UBRE=0.3608). Identical results occurred for FFT-B (UBRE=0.36048, deviance explained 51.7%, no significant effects): therefore, it must be stressed that SSA decomposition provides a better explanation of the data, the number of predictors entering the model being equal.

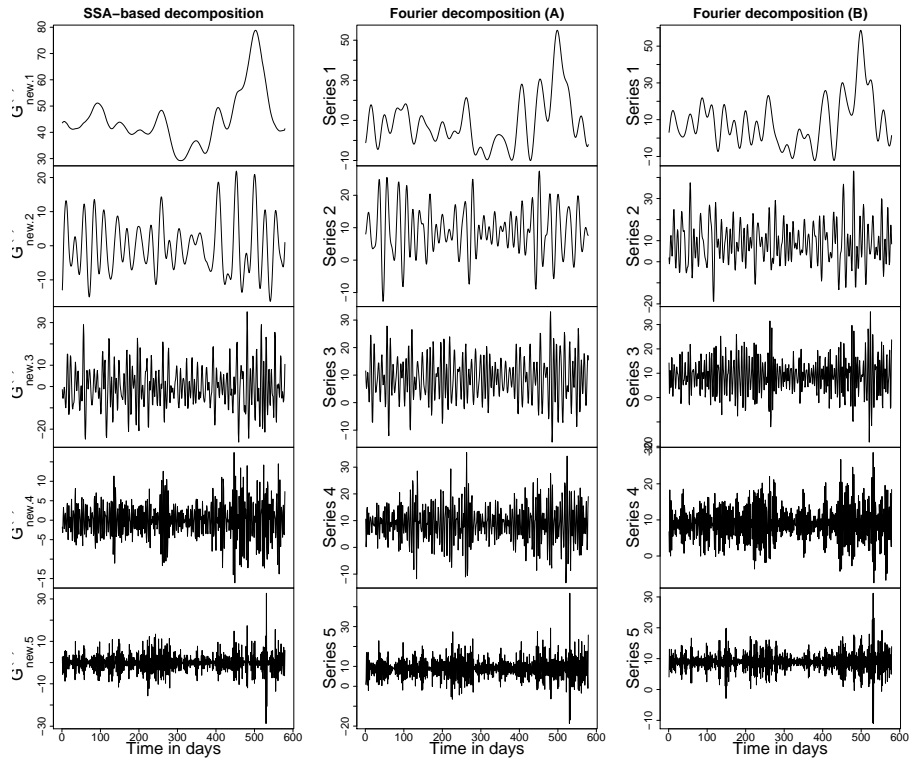

FIG 8. *LEFT: SSA-based decomposition according to the following grouping: $\mathcal{G}_{new.1} = \{1-2\}$, $\mathcal{G}_{new.2} = \{3-6\}$, $\mathcal{G}_{new.3} = \mathcal{G}_2$, $\mathcal{G}_{new.4} = \mathcal{G}_3$, $\mathcal{G}_{new.5} = \mathcal{G}_4$. MIDDLE: Fourier decomposition obtained by applying the R decompose() function with breaks equal to $1, 19, 41, 83, 165, 579$ days (except for the first and the last breaks, these were obtained as $579/r$, where $r = 30, 14, 7, 3.5$). RIGHT: Fourier decomposition obtained by setting breaks equal to $1, 21, 80, 146, 220, 579$ (except for the first and the last breaks, these were defined as $579/\Pi(\mathcal{G}_{new.j}^{(3)})$ for $j = 2, 3, 4, 5$).*

Most of the literature report little evidence for mortality displacement due to $PM_{10}$: our findings reinforce the conclusion that the increased chronic morbidity risks associated with even small increase in $PM_{10}$ are well established, although these increases are not greater for susceptible populations. In concordance with our results, several authors report a lag of about four weeks between the pollutant exposure and an increase of the mortality/morbidity risk: for example, [21] reported significant effects on cardiovascular-respiratory mortality in Sydney, Australia, at longer timescales (one month or more). Similarly, [36] assumed a distributed lag model for mortality due to natural causes in Milan, Italy, including lags up to 45 days: the authors reported an estimated total suspended particulate relative risk RR=1.037 (C.I.: 1.019-1.056) for the first 15 days, close to zero for lags between 16 and 30, and RR=1.027 (C.I.: 1.019-1.56) for lags up to 45 days.

TABLE 5
*Effect estimates, approximate significance of smooth terms and global model scores: a Fourier decomposition obtained by applying the `R decompose()` function was used, with breaks set equal to* $1, 19, 41, 83, 165, 579$ *days*

| Approximate significance of parameter estimates | | | | |
|---|---|---|---|---|
| | Estimate | Std. Error | z value | $Pr(>|z|)$ |
| (Intercept) | 2.1959872 | 0.0465239 | 47.201 | <2e-16 |
| pm10.1.new(2) | 0.0003945 | 0.0012542 | 0.315 | 0.753 |
| pm10.2.new(2) | -0.0003446 | 0.0016133 | -0.214 | 0.831 |
| pm10.3.new(2) | -0.0012409 | 0.0015482 | -0.802 | 0.423 |
| pm10.4.new(2) | -0.0012339 | 0.0015233 | -0.810 | 0.418 |
| pm10.5.new(2) | 0.0026280 | 0.0017240 | 1.524 | 0.127 |
| Approximate significance of smooth terms | | | | |
| | edf | Chi.sq | p-value | |
| S(day) | 8.881 | 187.811 | <2e-16 | |
| S(temp) | 2.214 | 9.840 | 0.0799 | |
| S(umr) | 1.888 | 5.054 | 0.2818 | |
| Global scores | | | | |
| | R-sq.(adj) | Dev. explained | UBRE score | n |
| | 0.502 | 51.7% | 0.3608 | 579 |

## 5. Discussion and conclusions

A great deal of uncertainty exists about the extent of life-shortening or the increase in morbidity due to the effects of air pollution. A limited amount of results from particulate matter time series studies suggests that the increased morbidity/mortality risk is not greater for susceptible populations. If a mortality/morbidity displacement (harvesting) effect is evident only a few days after exposure, then relevance of the findings of the daily time-series studies could be questioned, as adverse health effects might be arguably attributed only to the low quality of frail individuals at risk.

To test and estimate the pattern of mortality displacement, we proposed a statistical framework based on Singular Spectrum Analysis (SSA), a geometric technique derived from dynamical system theory, suitable for constructing easily interpretable exposure variables at several timescales. We believe that our method is superior from a practical point of view than FFT-based methods: the decomposition of the original series can be seen as a part of a data-driven process, and Fourier frequencies do not need to be fixed in advance. The only free parameter is the window length $L$. The problem at hand can suggest a sensible value, for the reason that the main (and only) rule of thumb for stationary series containing multiple harmonic frequencies prescribes that periods larger than $L$ are confused with long-term effects.

Promising theoretical developments are reported in [15], where SSA is extended to deal with missing data: the issue of missing information is ubiquitous in meteorologic and pollutant time-series, and the power of such newer data pre-processing methods has still to be established. Some subjective adjustments are still needed during the grouping phase. In particular, a correct separation of the dominant period (trend) from sub-harmonic frequencies merged with the sub-dominant group is a critical phase: we believe that the data-adaptive wavelet-based method introduced in [27] is the correct route towards a fully automated data analysis in multi-scale public health time-series studies. More efficient functional clustering algorithms will be needed too: for example, the shape-based curve clustering procedure described in [18] is very promising.

**Acknowledgements and additional information**

We thank Prof. Gabriella Serio and Dott. Rosi Prato, Apulian Regional Epidemiological Center, for the epidemiological data. We also thank Ing. Vincenzo Campanaro, Municipality of Bari, Department of Environmental Protection and Health, for data pollutant data. We would also like to thank two anonymous referees and the Associate Editor for their helpful comments. A special thank to Dott. Giusi Graziano, Apulian Regional Epidemiological Center, for her support.

A collection of R functions [24] for SSA and related stuff is available by the authors upon request. A suitable alternative is the SSA-MTM toolkit, freely available at http://www.atmos.ucla.edu/tcd/ssa/.